\documentclass[final,a4paper,11pt,3p,twocolumn,times]{elsarticle}

\usepackage{graphicx}
\usepackage{amsmath, amssymb, amsfonts}
\usepackage{hyperref}
\usepackage[american]{babel}

\begin{document}

\title{On The Period Determination of ASAS Eclipsing Binaries}

\author[itb]{L. Mayangsari\corref{cor1}}
\ead{lidiaisme@gmail.com}
\cortext[cor1]{Corresponding author}

\author[itb]{R. Priyatikanto}
\author[itb,bosscha]{M. Putra}

\address[itb]{Prodi Astronomi Institut Teknologi Bandung, Jl. Ganesha no. 10, Bandung,\\ Jawa Barat, Indonesia 40132}
\address[bosscha]{Bosscha Observatory, Lembang, Jawa Barat, Indonesia}

\journal{Proceeding of ICMNS 2012}

\begin{abstract}
Variable stars, or particularly eclipsing binaries, are very essential astronomical occurrence. Surveys are the backbone of astronomy, and many discoveries of variable stars are the results of surveys. All-Sky Automated Survey (ASAS) is one of the observing projects whose ultimate goal is photometric monitoring of variable stars. Since its first light in 1997, ASAS has collected 50,099 variable stars, with 11,076 eclipsing binaries among them. In the present work we focus on the period determination of the eclipsing binaries. Since the number of data points in each ASAS eclipsing binary light curve is sparse, period determination of any system is a not straightforward process. For 30 samples of such systems we compare the implementation of Lomb-Scargle algorithm which is an Fast Fourier Transform (FFT) basis and Phase Dispersion Minimization (PDM) method which is non-FFT basis to determine their period. It is demonstrated that PDM gives better performance at handling eclipsing detached (ED) systems whose variability are non-sinusoidal. More over, using semi-automatic recipes, we get better period solution and satisfactorily improve 53\% of the selected object's light curves, but failed against another 7\% of selected objects. In addition, we also highlight 4 interesting objects for further investigation.

\end{abstract}

\maketitle

\section{Introduction}

Variable stars, in particular eclipsing binaries, are very essential astronomical object to open our knowledge on stellar properties. Eclipsing binary observations provide physical constraints to construct the theory of stellar physics and evolutions. Despite these advances, 90\% of all bright variable stars ($m<12$) have not been discovered \citep{pacz}. in order to collect more data on variable stars, astronomers have initiated astronomical survey projects, for surveys constitute the backbone of astronomy and many breakthroughs have been achieved from them.

One of the latest sky survey projects is ASAS \citep{poj97,poj09}. The program started in 1996 using a small instrument that automatically operated to obtain photometric data and variability. Since its first light in 1997, ASAS has collected 50,099 variable stars, with 11,076 of them are classified as eclipsing binary systems \citep{pacz}.

As a survey observation, ASAS certainly does not warrant any star observed with the same frequency. ASAS photometric data unevenly distributed in the time domain, so the determination of the period of star variability becomes more difficult. Analysis of Variance (AoV, \citep{czerny}) method adopted in ASAS does not always provide a good period solution with good light curve.

Therefore, we focused on the process of period determination of eclipsing binaries in ASAS catalog occupying other methods. The methods used in the present study are the Lomb-Scargle algorithm \citep{lomb,scargle} and Phase Dispersion Minimization \citep{stellingwerf}. Additional recipes combined with the period determination of the basic methods are used to obtain the best period solution.

\begin{figure*}[t]
\centering
\includegraphics[width=0.75\textwidth]{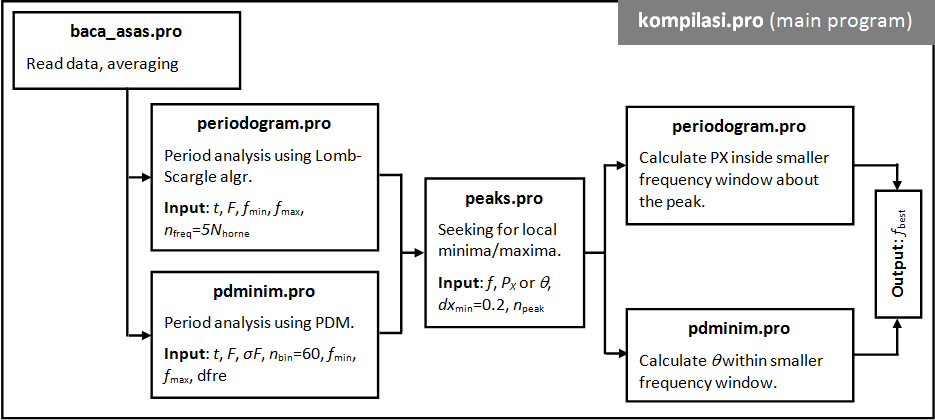}
\caption{Schematic diagram of the implemented recipes to determine the most suitable period.}
\label{chart}
\end{figure*}

\section{Data}
ASAS (ASAS-3N, ASAS-3S) uses a CCD camera with dimension of $2048\times2048$ pixels, 15 microns each pixel. The camera is combined with a telephoto lens diameter of $D \sim 10$ cm and  focal length of $f\sim20$ cm. This configuration can capture the sky area of $\sim$4 square degrees. Using I-band (880 nm) and V-band (550 nm), ASAS can detect bright stars with $V\lesssim13$.

After 2-3 minutes exposure, each frame will undergo standard data reduction procedures and photometric analysis using asymmetric point-spread function (PSF) \citep{poj97}. Astrometry analysis (determination of stars' coordinates) is done by triangulation, based on the stars in the Guide Star Catalog (GSC). These steps will culminate in the process of writing data to astrometry and photometry catalog. ASAS photometry catalog contains the identity of the stars, the positions (RA-dec), and five visual magnitudes which are taken with different exposure times. ASAS photometric data has a grade that determined according to appearing photometric error.

By \citep{pacz}, the objects' periodicity are analyzed and classified into three main groups, namely EC (eclipsing contact), ESD (eclipsing semi-detached), and ED (eclipsing detached) by its Fourier coefficients \citep{rucinski}. Based on this catalog of eclipsing binary stars we chose several objects for further analysis. The selection of objects is based on the position and magnitude of the objects that can be observed at Bosscha Observatory, Indonesia ($107.6^{\circ}$ East and $6.8^{\circ}$ South). Here are the criteria used for objects selection:
\begin{description}
\item\textbf{RA:} right ascension between 12 to 24 hours.
\item\textbf{dec:} located at declination of $-60^{\circ}$ to $30^{\circ}$.
\item\textbf{magnitude:} visual magnitude at primary minimum (VMAX) between 8.5 to 9.5. Theses limits are determined by the saturation limit and limiting magnitude of small telescopes at Bosscha Observatory.
\item\textbf{number of valid data ($N_0$):} selected objects have about 500 valid photometric data.
\end{description}

This selection process produced 149 eclipsing binaries. In this study, we only select 30 of them, 10 objects for each type. For each selected objects, we evaluate the quality of the light curve and provide grades:
\begin{description}
\item\textbf{grade 1:} the light curve is sharp and smooth
\item\textbf{grade 2:} profile seem to be obvious, but there is a spread in the time domain (period is less precisely determined)
\item\textbf{grade 3:} the profile and the period is not clear
\end{description}

\section{Methods}
Before undergoing periodicity analysis, we calculate the averaged magnitude of each star in each observation using
\begin{align}
\bar{m}&=m_0-2.5\log{\bar{f}}\\
\bar{f}&=\dfrac{1}{5}\sum_{i=0}^{4}10^{0.4(m)0-m_i)}
\end{align}
where $m_0$ represents magnitude of one exposure, $m_i$ magnitude at other exposures, while $\bar{f}$ is the average value of relative flux.

This calculation will give an averaged magnitude error of:
\begin{align}
\bar{\sigma_m}&=\sqrt{\sigma_0^2}+\left(\dfrac{2.5}{\ln{10}}\dfrac{\bar{\sigma_f}}{\bar{f}}\right)\\
\bar{\sigma_f}&=\dfrac{1}{5}\sqrt{\sum_{i=0}^{4}\left[\left(\dfrac{\ln10}{2.5}f_i\right)^2(\sigma_0^2+\sigma_i^2)\right]}
\end{align}
where $\sigma_0$ is magnitude error of one exposure while $\sigma_i$ represent magnitude error in another exposures.

\subsection{Lomb-Scargle Algorithm}
Analysis of periodicity is done using Lomb-Scargle algorithm \citep{lomb,scargle} which is implemented in the IDL language program (scargle.pro). The program requests several inputs, which are photometric data $m_i$, time $t_i$ and the constraints of tested period or frequency. At each tested frequency, a statistical parameter $P_N(\omega)$ will be calculated using fast algorithm as defined by \citep{press}. The value $P_N(\omega)$ corresponds to the appropriate period which will yield a good light curve.

\begin{figure*}[ht]
\centering
\includegraphics[width=0.8\textwidth]{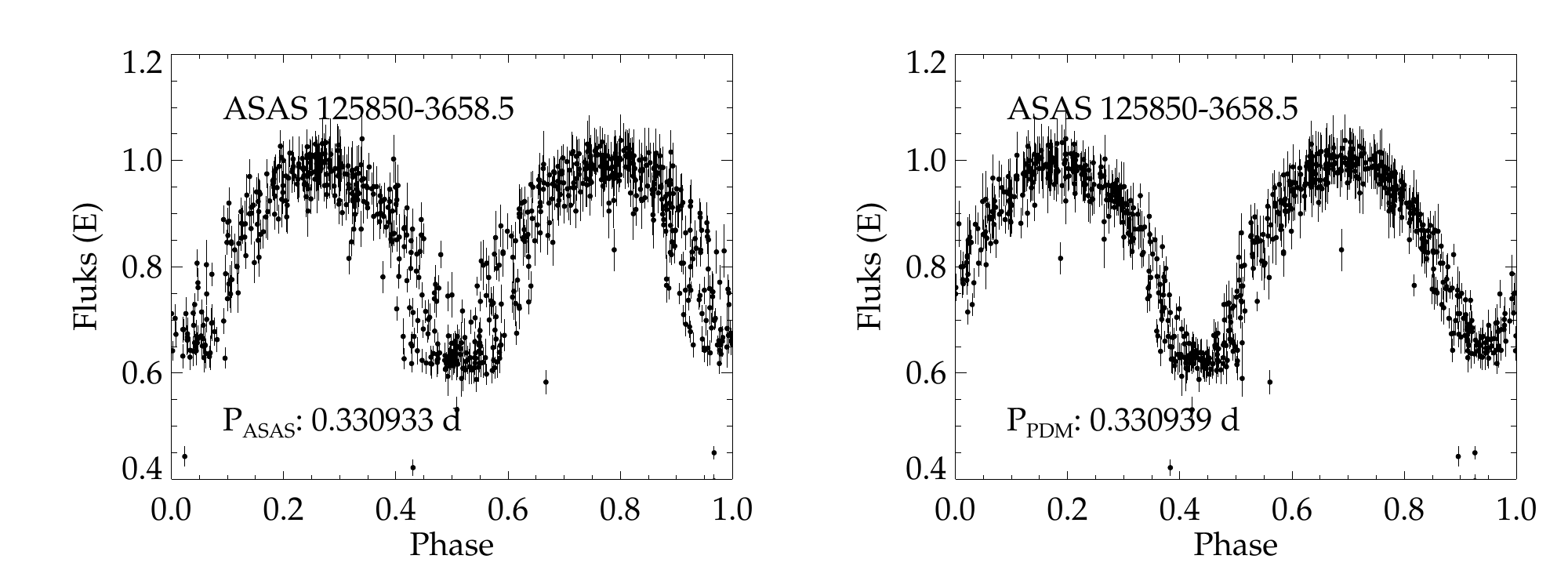}
\caption{Light curve of ASAS125850-3658.5 eclipsing contact binary folded using catalogued period by ASAS' team (left) and the best solution from this study (right). It is shown that our analysis slightly improve the light curve.}
\label{compare}
\end{figure*}

\subsection{Phase Dispersion Minimization}
Analysis of periodicity using PDM \citep{stellingwerf} is performed using IDL program (pdm.pro) written by Mark Buie. Almost the same as the previous program, this program receives inputs of magnitude $m_i$, error $\sigma_i$, as well as the observation time $t_i$. Like other periodicity analysis method, statistical parameter calculations are performed for the specified number of tested frequency. Statistical parameters $\theta_{PDM}$ will give a minimum value when the tested frequency corresponds to the frequency of the object.

In the analysis of the periodicity using PDM, subharmonic frequencies that are simply multiplication of the correct frequency may have also low statistical value. This phenomenon can obscure the solution of periodicity analysis.

\subsection{Additional Recipes}
In our study, two methods that have been mentioned do not work straightforwardly, but rather work with particular recipes. There are several important parts of the recipe that we use. The scheme of this recipe can be seen in Figure 1.

The first is the number and the range of tested frequencies. The number of frequencies that are being analyzed is $5N_{freq}$, where $N_{freq}$ is the optimal value by \citep{horne},
\begin{equation}
N_{freq}=-6.362+1.193N_0+0.00098N_0^2
\end{equation}

Tested frequencies or periods are within a certain range that appropriate to the type of object to be explored, ie, between 0.15 days to 150 days. To speed up the process, the range is divided into two classes, namely 0.15 - 15 and 15 - 150. Among this range, the period analysis will give several frequencies with extreme statistical values (maximum for Lomb-Scargle and minimum for PDM).

The second part is the selection of a local extreme points. Both adopted methods used in the analysis will give a number of local maximum points that do not always coincide with the best solution. Frequency resolution, leakage \citep{scargle} and subharmonic \citep{stellingwerf} may be the cause. Therefore, we select top-10 spots to be examined.

The third part is the study of 10 chosen frequencies with narrower frequency window, which is 0.1 times the width of the initial window. In this second stage of analysis, global extreme point (fgood) will be selected as the best solution for any window. Then, photometric data will be folded with this frequency and its light curve quality will be assessed. This assessment is done by manually observing the distribution of the data in the formed light curve. Best light curve corresponds to the solution frequency/period.

\section{Results}
After passing the period analysis, the 30 selected objects have the best light curves with their final frequency or period. Each method will be compared based on the quality of the resulting light curves. This assessment is the same as before, but coupled with a comparative assessment.
\begin{description}
\item\textbf{Grade A}: if the resulting light curve is better than ASAS light curve
\item\textbf{Grade B}: if the quality of the light curve remains
\item\textbf{Grade C}: if the resulting light curve is worse than ASAS light curve
\end{description}

\subsection{Comparison of the Results}
Furthermore, the results of each method of analysis will be compared to each other. \autoref{tabel1} and \autoref{tabel2} summarize both Lomb-Scargle algorithm and PDM performance.

\begin{table}[ht]
\centering
\caption{Statistics of the light curves obtained using LS.}
\label{tabel1}
\begin{tabular}{cccc}
\hline
 & EC & ESD & ED \\
\hline
A & 6 & 3 & 3 \\
B & 2 & 6 & 3 \\
C & 2 & 1 & 4 \\
\hline
\end{tabular}
\end{table}

\begin{table}[ht]
\centering
\caption{Statistics of the light curves obtained using PDM.}
\label{tabel2}
\begin{tabular}{cccc}
\hline
& EC & ESD & ED \\
\hline
A & 7 & 5 & 4 \\
B & 3 & 4 & 5 \\
C & 0 & 1 & 1 \\
\hline
\end{tabular}
\end{table}

From these two tables, it can be seen that the success rate of the two methods is good and comparable enough to examine eclipsing binaries type EC, especially PDM method (see \autoref{compare}). Variability of EC type eclipsing binaries had a brief period. The wide minima and amplitude variability is large enough to produce statistical value, $P_N$ and $\theta_{PDM}$, that contrast one and another. Lomb-Scargle algorithm performance deteriorates when dealing with ED type eclipsing binaries whose a long period, narrow minima, and the light curve character that are far from initial sinusoidal function. Lomb-Scargle algorithm which work based on fitting a sinusoidal function to data often fail to get the best frequency solutions \citep{stellingwerf}. All results is summarized in \autoref{hasil}.

\vspace{10pt}
\begin{table*}[ht]
\centering
\caption{Period analysis of 30 ASAS eclipsing binary stars performed in this study. The last column shows the method used to obtain the best light curves.}
\label{hasil}
\begin{tabular}{cccc}
\hline
ASAS-ID & $P$ (day) & Gr. & Remark \\
\hline
212100-5228.7  -- ED & 0.435333 & 1 & PDM \\
205441-4543.9  -- ED & 0.986360 & 1 & ASAS \\
193833+1706.5  -- ED & 1.474913 & 1 & PDM, LS \\
201310+1020.7  -- ED & 1.705476 & 1 & PDM, LS \\
173923-3408.4  -- ED & 2.147984 & 1 & PDM \\
203943+0323.6  -- ED & 3.502981 & 2 & PDM \\
183458+1126.6  -- ED & 4.553585 & 1 & PDM \\
173708-4036.7  -- ED & 5.088687 & 1 & PDM, LS \\
180951-1533.0  -- ED & 28.352129 & 3 & PDM \\
125850-3658.5  -- EC & 0.330938 & 1 & PDM \\
143504+0906.8  -- EC & 0.355158 & 1 & PDM \\
144619-3541.7  -- EC & 0.365944 & 1 & PDM, LS \\
182913+0647.3  -- EC & 0.375308 & 1 & PDM \\
125340-5010.6  -- EC & 0.404663 & 1 & PDM \\
180921+0909.1  -- EC & 0.409009 & 1 & PDM \\
204539-5102.7  -- EC & 0.453644 & 1 & PDM, LS \\
234535+2528.3  -- EC & 0.578458 & 1 & PDM \\
160017-4507.6  -- EC & 0.647870 & 1 & PDM, LS \\
203937+1425.7  -- EC & 0.844652 & 1 & PDM, LS \\
173758-3911.4  -- ESD & 0.303315 & 1 & PDM, LS \\
133138-5146.3  -- ESD & 0.738454 & 1 & PDM, LS \\
221139+2607.5  -- ESD & 1.072820 & 1 & ASAS \\
213126+2627.7  -- ESD & 1.510180 & 1 & PDM, LS \\
132521-5946.9  -- ESD & 1.874696 & 1 & PDM \\
153319-2501.6  -- ESD & 2.267258 & 2 & PDM \\
160457-4617.6  -- ESD & 2.556834 & 2 & PDM \\
190410-0201.8  -- ESD & 2.733943 & 1 & PDM, LS \\
181847-1817.5  -- ESD & 3.338764 & 1 & PDM, LS \\
120924-4525.6  -- ESD & 64.668778 & 3 & PDM \\
\hline
\end{tabular}
\end{table*}

\subsection{Interesting Objects}
As an extra results of this study, we find some interesting objects to be studied/observed more deeply.
\begin{description}
\item[ASAS22503-5348.6] which has little variability, narrow minima, and there is not much data around the minima. As a result, the analysis of periodicity that we do provide solutions with period that is much different than that of the ASAS catalog.
\item[ASAS190410 and ASAS203937] are two eclipsing binaries typed ESD and EC with a skewed minima that may have a third object. O-C analysis needs to be done to confirm this hypothesis.
\item[ASAS160457-4617.6] is an eclipsing binaries typed ESD with an unclear period solution. Solutions conducted by either ASAS, Lomb-Scargle, or PDM do not result in the best light curve. There is still a spread in the phase domain that can be caused by the lack of proper solutions period or a change in the period.
\end{description}

\section{Conclusions}
We have conducted an analysis of the 30 samples of ASAS eclipsing binaries using Lomb-Scargle algorithm and Phase Dispersion Minimization which are implemented with additional recipes. Compared to ASAS catalogue, our period analysis can satisfactorily improve 53\% of the selected object's light curves, but failed against another 7\% of selected objects. Lomb-Scargle algorithm is not so good at handling ED type eclipsing binaries. In addition, we found 4 objects with unusual light curves for further investigation.

\newcommand{\apj}{Astroph. Journ.}
\newcommand{\mnras}{Mon. Not. Royal Astron. Soc.}
\newcommand{\apjs}{Astroph. Journ. Supp.}

\bibliographystyle{plainnat}
\bibliography{paper-asas}

\begin{thebibliography}{10}
\providecommand{\natexlab}[1]{#1}
\providecommand{\url}[1]{\texttt{#1}}
\expandafter\ifx\csname urlstyle\endcsname\relax
  \providecommand{\doi}[1]{doi: #1}\else
  \providecommand{\doi}{doi: \begingroup \urlstyle{rm}\Url}\fi

\bibitem[Horne and Baliunas(1986)]{horne}
J.~H. Horne and S.~L. Baliunas.
\newblock {A Prescription for Period Analysis of Unevenly Sampled Time Series}.
\newblock \emph{\apj}, 302:\penalty0 757, March 1986.

\bibitem[Lomb(1976)]{lomb}
N.~R. Lomb.
\newblock {Least-square Frequency Analysis of Unequaly Spaced Data}.
\newblock \emph{\apjs}, 39:\penalty0 447, June 1976.

\bibitem[Paczy{\'n}ski et~al.(2006)Paczy{\'n}ski, Szczygiel, Pilecki, and
  Pojma{\'n}ski]{pacz}
B.~Paczy{\'n}ski, D.~M. Szczygiel, B.~Pilecki, and G.~Pojma{\'n}ski.
\newblock {Eclipsing Binaries in the All Sky Automated Survey Catalogue}.
\newblock \emph{\mnras}, 368:\penalty0 1311, December 2006.

\bibitem[Pojmanski(1997)]{poj97}
G.~Pojmanski.
\newblock {The All Sky Automated Survey}.
\newblock \emph{Acta Astron.}, 47:\penalty0 467, December 1997.

\bibitem[Pojma{\'n}ski(2009)]{poj09}
Grzegorz Pojma{\'n}ski.
\newblock {The All Sky Automated Survey (ASAS) and Bohdan Paczy{\'n}ski's Idea
  of Astronomy with Small Telescopes}.
\newblock In Krzysztof~Z. Stanek, editor, \emph{The Variable Universe: A
  Celebration of Bohdan Paczy{\'n}ski}, volume 403. ASP Conference Series,
  December 2009.

\bibitem[Press and Rybicki(1989)]{press}
W.~H. Press and G.~B. Rybicki.
\newblock {Fast Algorithm for Spectral Analysis of Unevenly Sampled Data}.
\newblock \emph{\apj}, 338:\penalty0 227, March 1989.

\bibitem[Rucinski(1997)]{rucinski}
S.~M. Rucinski.
\newblock {Eclipsing }.
\newblock \emph{Astron. Journ.}, 113:\penalty0 1112, December 1997.

\bibitem[Scargle(1982)]{scargle}
J.~D. Scargle.
\newblock {Studies in Astronomical Time Series Analysis. II. Statistical
  Aspects of Spectral Analysis of Unevenly Spaced Data}.
\newblock \emph{\apj}, 263:\penalty0 835, December 1982.

\bibitem[{Schwarzenberg-Czerny}(1989)]{czerny}
A.~{Schwarzenberg-Czerny}.
\newblock {On the Advantage of Using Analysis of Variance Period Search}.
\newblock \emph{\mnras}, 241:\penalty0 153, January 1989.

\bibitem[Stellingwerf(1978)]{stellingwerf}
R.~F. Stellingwerf.
\newblock {Period Determination Using {Phase Dispersion Minimization}}.
\newblock \emph{\apj}, 224:\penalty0 953, September 1978.

\end{thebibliography}

\end{document}